\documentstyle[preprint,revtex]{aps}

\begin{document}

\begin{title}

Electron focusing, mode spectroscopy and mass enhancement\\
in  small GaAs/AlGaAs rings
\end{title}

\author{J.\ Liu$^1$, K.\ Ismail$^{2,4}$, K.\ Y.\ Lee$^{3,4}$, J.\ M.\ Hong$^4$
\\
and S.\ Washburn$^1$}

\begin{instit}

$^1$ Department of Physics and Astronomy, The University of North Carolina,\\
Chapel Hill, NC 27599-3255, USA.\\
$^2$ Dept.\ of Telecommunication and Elec.\ Eng., University of Cairo,\\
Cairo, Egypt.\\
$^3$ IBM East Fishkill, Hopewell Junction, NY 12533, USA.\\
$^4$ IBM Watson Research Center, Yorktown, NY 10598, USA.
\end{instit}



\begin{abstract}
\vspace{-36pt}

A new electron focusing effect has been discovered in small single and coupled
GaAs/AlGaAs rings.  The
focusing in the single ring is attributed solely to internal orbits.  The
focusing effect allows the ring to be used as a small mass
spectrometer.  The focusing causes peaks in the magnetoresistance at low
fields,
and the peak positions were used to study the dispersion relation
of the one-dimensional magnetoelectric subbands.  The electron effective mass
increases with the applied magnetic field by a factor of $50$, at a magnetic
field of $0.5T$.  This is the first time this increase has been measured
directly.  General agreement obtains between the experiment and the subband
calculations for straight channels.

\vspace{15pt}
72.20.My, 73.20.Dx, 85.30.Tv.\\
\today,  submitted to Physical Review Letters

\end{abstract}
\newpage

In 1897, Thompson\cite{Kelvin} invented the mass spectrometer, which was later
refined by
Bainbridge\cite{Bainbridge} and others.  Mass spectroscopy involves
bending the trajectory of particles emitted from a point source with a magnetic
field and using a point collector to select a particular orbit size, $l_c =
mv/eB$.  Electron focusing
in the solid state is analogous to this magnetic focusing of electrons in
vacuum.  A magnetic field is used again to focus the electrons injected by a
point
contact into a second point contact a distance $L$ away\cite{sharvin},
but in the solid state, the electrons can bounce specularly from walls in the
sample\cite{tsoi,HvH}.  In the conventional mass spectrometer, the momentum of
the particle is fixed by an external electric field that accelerates the
particle, but in the solid states experiments, the Fermi energy determines the
velocity\cite{sharvin}.  The focusing condition
is that the cyclotron orbit diameter 2$l_c$ be an integer fraction of $L$,
$L= 2jl_c = 2j\hbar k_F /eB$.  Focusing
peaks appear in the magnetoresistance R(B) because of skipping orbits such
as the external trajectories on the two-ring sample in the inset of Figure 1.
Here $l_c$ is the cyclotron radius, and $j-1 = 0,1,2,3...$ is the number of
bounces in the trajectory.  Solid state focusing has been employed to study
surface scattering in metals\cite{tsoi}, and electron propagation in
semiconductors\cite{HvH}.  In these experiments, the electron
traveled
in the half space bounded by the surface through which they were injected,
and they bounced specularly from that smooth surface, but focusing can
occur for electrons confined to move in circular tracks\cite{Bainbridge} such
as
the loop depicted
in the inset of Fig.~1\cite{washburn91}.  In such a loop, electrons inside
the ring are focused
away from the outlet port when $l_{c}$ matches the loop radius r.  As a result,
the ring acts like a trap: The resistance is enhanced and a focusing peak
appears because of the internal orbit in the inset of Figure~1.  If $n$
transvese modes
(subbands) are occupied in the wires forming the loop, then $n$ different
values of velocity parallel to the wire can be focused to achieve $n$
peaks in R(B) as each $l_{cn}$ matches r.
This allows the loop to be used as a masss spectrometer that measures both
energy and
tangential momentum, and so map the dispersion relation $E_n (k_n ,B)$ for the
modes in the wire.  As we will show this allows us to make the first
measurements of the enhancement of the cyclotron effective mass as the
magnetic field increases and the dispersion in the
magnetoelectric subbands\cite{berggren,wees}.

Our samples were fabricated on a standard high-mobility GaAs/AlGaAs
modulation-doped layer (carrier density $n_s = 2 \times 10^{15}/m^2$ and
mobility $\mu = 90 m^2/Vsec$) grown by molecular beam epitaxy.  The ring
geometry was defined by a shallow mesa formed through a wet etching
technique\cite{washburn91}, and metal gate covers all active portions of the
device.  More details of the fabrication are reported
elsewhere\cite{washburn91}.
The rings have radii $r = 0.85 \mu m$ and line widths $t = 0.3 \mu m$.
Earlier experiments on similar samples have shown clear conductance
quantization
and electrical focusing effects at $T = 4.2K$, and large Aharonov-Bohm
oscillations up to a (20\% modulation of R(0)) at $T~<
0.1K$\cite{washburn91}.  This was the first time that the mode counting effects
were seen at such length scales, and the first observation of focusing in
loops.
Such evidence assures us that the transport in the loops is ballistic.
All experiments reported here were conducted at $T = 4.2K$.
The resistance was measured with a standard four-probe lock-in method
at a frequency of $23.6Hz$ and at $1mV$ (rms) excitation voltage.

Figure~1 shows the conductance $G$ as a function of gate voltage $V_g$ at
$B=0$
for the single ring and double rings.  Mode counting conductance
steps\cite{steps,beenakker91} occur in both single and double ring samples.
The
average heights of the
steps are $0.75$ and $1.3\ e^2/h$, respectively, which are consistent with the
results of the previous experiment\cite{washburn91}.  In this figure we also
show where the subband bottoms $E_n (0,0)$ cross $E_F$ in the
single ring, as determined by the peaks in derivative $dG/dV_g$.

Figure 2 contains R(B) for the single ring, which contains a series of peaks on
a smooth background.  (We remark that similar peaks have been observed
previously in rings formed
on samples patterned without any etching.\cite{ford})  Analysis of the double
ring data is
complicated because of the competition between internal and external orbits,
so we will discuss only the single ring, where focusing results entirely from
internal orbits, from now on.

We are interested in the focusing peaks, so
we have subtracted the smooth part of R(B).
We are not
sure of the origin of the large negative R(B) in our
experiment, but we speculate that it might result in part from steering the
incoming
beam away from head-on collision with (and concommitant strong reflection from)
the inner wall of the loop.
We used a quasi-Lorentzian functional form,
\begin{equation}
R(B) = R_0 - \frac{\Delta}{1 + (B/B_c)^{2.3}}
\end{equation}
to fit R(B) (dashed line).
The power $2.3$ is chosen simply because it gives us the best fit, but the
results discussed below do {\em not} depend on this choice.
Since the focusing effect at the outlet port increases the resistance,
we only used the low resistance portions (heavy lines) in the fit,
allowing both $\Delta$ and $B_c$ as free parameters.
After subtracting the smooth part, we
nominally have the net contribution of the
focusing, which contains several peaks.
In addition, one sees the onset of Shubnikov-deHaas (SdH) oscillations ($\sim
0.5 T$ to $\sim 1T$).  The peak positions are then mapped
by fitting a Gaussian function,
\begin{equation}
\Delta R = Ie^{- [ {(B-B_0) /\Delta B} ]^2} +aB+c.
\end{equation}
to the data in the neighborhood of each peak, where the last two terms allow
for
imperfect subtraction and overlap of the peaks.  The resulting center field
$B_0$, normalization factor $I$, and mean deviation $\Delta B$ yield the peak
position, intensity and width, respectively.  We will discuss the latter two
parameters more in a forthcoming publication, and here we will concentrate
interpreting the peak positions.

Starting with the relation $l_c =\hbar k_F /eB$, one can show that the
cyclotron
effective mass is $m_{cn} = \hbar^2 k/(dE_n /dk)$ and tangential (group)
velocity is $v_{n} = dE_n/\hbar dk$\cite{dispersion}.  The
focusing condition now becomes $r = l_{cn} = \hbar k_{Fn}/eB\ $, which is the
same form as in free space, but with the
Fermi wave-vector $k_{Fn}$ {\em tangential to the arm of the loop}.
This allows us to infer $k_{Fn}$ from the peak positions.
After measuring $E_F (V_g )$
via the SdH effect ($0<B<6T$) to be $E_F = 5.98V_g + 15.4\ (meV)$, we have
the dispersion relation directly.  Figure~3 contains $E_F$ as a function of
peak
position $B_0$ (upper axis) and (through the focusing condition) the
corresponding wave-vectors (lower axis) in convenient units $2\pi/r$.  Data
from
$n=1$ is omitted because it is sensitive to the functional form (1).

To our knowledge, there have been no calculations of the dispersion
relation of a circular channel, but there have been quite a few
discussions about the straight channels\cite{beenakker91}.
Since for the data of interest, the ring radius is at least two orders
of magnitude larger than the Fermi wavelength (notice the order of
magnitude on the lower abscissa for Fig.~3), we will use the results
for straight channels.
A good model of the electrostatic confinement potential
comprises a flat bottom plus parabolic walls\cite{laux86,berggren},
$V(|x|>t/2) = m \omega_0^2(|x| - t/2)^2/2$ and $V(|x|<t/2)=0$,
where $m=0.067m_e$ is
the effective mass for two-dimensional GaAs/AlGaAs (100) electron gas.
At $B=0$, the problem has been solved analytically\cite{poole85} with the
result
\begin{equation}
E_n(k,0) = E_n(0,0) + \frac {\hbar ^2 k^2}{2m} ,
\end{equation}
where the offset of the subband bottom is
\begin{equation}
E_n(0,0) = \left \{ E_{P} \left [ \frac {4(n + 1/2 )}{E_{P}} +
\frac {1}{E_{S}} \right ] ^ {1/2}
- \left [ \frac {1}{E_{S}} \right ] ^ {1/2}
\right \} ^ 2
\end{equation}
where $E_{S} = (\pi\hbar)^2/2mt^2$ and $E_{P}=\hbar \omega _0 /2$ are
ground state energies for square (width t) and parabolic wells,
respectively.
The effective mass is not changed by the electrostatic confinement.

We can now compare the measured $E_n(0,0)$ from Figure~1 ($B=0$) with these
formulae as shown in the inset of
Figure~4.  The agreement of the data with a parabolic form (solid line)
means that the deviation from a square well is
negligible.  The inferred well width is $t=0.3\mu m$, in
agreement with the lithographic linewidth.  The fit also yielded a barrier
energy $E_c = 13.1 meV$, which defines the bottom of the potential
well\cite{wees}.

For $B>0$, the electron is in a new potential
comprising the electrostatic part plus a magnetic potential
$ V_b= m \omega_c^2 (x-x_0)^2/2$, where $x_0 = -\hbar k/eB$.
For $t=0$,
the new potential is just a shifted parabola, so the
solution is a new harmonic oscillator\cite{kaplan,beenakker91},
but with a new curvature
$ \omega = (\omega_c^2 + \omega_0^2)^{1/2}$,
and an enhanced effective
$m^* = m \omega_c^2/\omega_0^2$.
The subband bottom rises
and eventually approaches the Landau level
when $\omega _c >> \omega_0$.  The dispersion relation is still
parabolic in $k$, but with a mass that increases with
B but is independent of n.

The solution for $B > 0$ and $t > 0$
is complicated by the flat part.\cite{berggren}
An effective mass $m^*$ still describes
the dispersion curve $ E_n(k,B) = E_{noff} + \hbar ^2  k^2/2m^*$,
with $m^*$ defined above, in the low B and high n limit.
Since this equation is not necessarily
true in the limit $k \rightarrow 0$, the significance of $E_{noff}$ is not
clear\cite{berggren}.  One must apply these formulae to the data advisedly
since B is not fixed while the dispersion in k is traced, but since the range
over which each peak is studied is small we can safely approximate B as a
constant in the analysis. The solid
curves in Fig.~3 are fits to (3) with $m^*$ and $E_{noff}$ as free parameters.
The resulting effective masses are displayed in Fig. 4.
The effective mass increases about $50$ fold!
For $n=7$ to $n=4$, $m^*$ is described quite well by the parabolic dependence
discussed above, but there is significant deviation at higher field where the
Landau level is beginning to form.
{}From the fit, we get $m^*(B=0)= 0.95$,
in excellent agreement with the ideal value $1$.

The enhancement of effective mass by the magnetic field in one-dimensional
subbands has been known theoretically for some time, but for two reasons, this
is the first experimental measurement.
First,
in the usual ballistic one-dimensional conductance
experiments,\cite{beenakker91} $k$ and $m^*$ (the density of states) cancel
each other and lead to the quantized conductance steps.
Second,
experiments designed to study the magnetoelectric
subbands, usually through the SdH effect,
measure the magnetic depopulation of the subbands,
and are intrinsically mostly sensitive to the subband bottoms. Except
in studies of the
sharpening of the depopulation peaks\cite{berggren},
little attention has been paid to the mass enhancement.
In contrast, our sample acts like a small mass spectrometer and provides a
direct measurement of the subband dispersion relation and the carrier mass.

The formula for the cyclotron orbit size is very general; it holds
even when the dispersion relation is not parabolic and
the effective mass $m^*$ loses its meaning.  Hence,
there is no particular reason we could not extend it into the
dispersionless range.
This point is worthy of
further experiments. That the transverse momenta of the modes are not
relaxed in the outlet region is interesting, too.  Potentially, it could
be used to study the mode relaxation and coupling.  Possible
variations, such
as widening the outlet, or placing a barrier in the outlet or in one arm of the
ring, could be designed.

In conclusion, we remark that we have observed the electron focusing effect in
a single ring.  The magnetoresistance peak positions, corresponding to
the cyclotron radii of different subbands matching the ring radius,
directly yielded the dispersion relation of the subbands.
Calculations for subbands in straight channels
agreed with the data if the lateral electric confinement was
modeled as a flat bottom plus a parabolic wall.
The fitted well parameters were self-consistent, and agreed
with other independent estimates.  The modeling allowed us to observe directly
the subband dispersion $E_n(k)$ and the enhancement of the cyclotron effective
mass.

ACKNOWLEDGMENT: This work was supported
with funding from IBM, the Microelectronics Center of
North Carolina and the University of North Carolina at Chapel Hill. We would
like to thank K. Li, V. Long, Y. Wang, and W. Gao for their assistance in
this experiment and M. B\"{u}ttiker, J. Davies and H. Baranger for valuable
conversations.

\newpage

\figure{The inset is a schematic for the single ring and coupled
double ring. The sizes marked are in the units of $\mu m$. The dashed lines
illustrate internal and external focusing orbits.
Conductance versus gate voltage is shown
for the single
ring and double rings. $T= 4.2K$. the position
of the turn-on voltage for the modes are
determined by the derivatives $dG/dV_g$ (dashed line) for the single ring.
}
\figure{R(B) the single ring at $V_g=0.55V$ (solid line), is fit to (1) (dashed
line), which accounts for the smooth background magnetoresistance.  The
difference between the two, $\delta R$ is the contribution of the focusing
effect.
}
\figure{The focusing peak positions
for subband indices 2 - 7.
The solid curves are fits to (11) treating
the effective mass $m^*$ and offset energy $E_{noff}$ as free
parameters.
}

\figure{The effective mass $m^*/m$ obtained from Figure~3 as
a function of the average magnetic field for each subband.  At low
field, it fits well with a parabolic function (solid line).
The inset contains the zero field subband bottom energy as a function
of the band indices $n$. The solid curve is a least-square fit to
the square well prediction.
}

\end{document}